\documentclass[aps,prd,onecolumn,superscriptaddress,showpacs,floatfix]{revtex4}
\usepackage{graphicx}
\usepackage{epsfig}

\begin{document}

\title{Equivalence between domain-walls and ''noncommutative'' two-sheeted
spacetimes: Model-independent matter swapping between branes}

\author{Micha\"{e}l Sarrazin\footnote{Also at Laboratoire de Physique du Solide (LPS), FUNDP}}
\email{michael.sarrazin@fundp.ac.be} \affiliation{Groupe
d'Application des MAth\'ematiques aux Sciences du COsmos
(GAMASCO),\\ University of Namur (FUNDP),
\\61 rue de Bruxelles, B-5000 Namur, Belgium}

\author{Fabrice Petit}
\email{f.petit@bcrc.be} \affiliation{Belgian Ceramic Research
Centre,\\4 avenue du gouverneur Cornez, B-7000 Mons, Belgium}

\begin{abstract}
We report a mathematical equivalence between certain models of universe
relying on domain-walls and noncommutative geometries. It is shown that a
two-brane world made of two domain-walls can be seen as a ''noncommutative''
two-sheeted spacetime under certain assumptions. This equivalence also
implies a model-independent phenomenology which is presently studied. Matter
swapping between the two branes (or sheets) is predicted through fermionic
oscillations induced by magnetic vector potentials. This phenomenon, which
might be experimentally studied, could reveal the existence of extra
dimensions in a new and accessible way.
\end{abstract}

\pacs{11.25.Wx, 11.27.+d, 02.40.Gh, 13.40.-f}

\maketitle

%
%
%

\section{Introduction}

During the last two decades, the possibility that our observable $(3+1)$%
-dimensional universe could be a sheet (a brane) embedded in a
higher-dimensional bulk spacetime has received a lot of attention. This line
of thought has shown to provide nice explanations to several puzzling
phenomena such as the hierarchy between the electroweak and the Planck
scales \cite{1}, the dark matter origin \cite{2} or the cosmic acceleration
\cite{3}. Domain-wall was demonstrated as a believable mechanism to explain
the trapping of the Standard Model (SM) particles on branes \cite{4},
especially fermions \cite{5,6,7,8,9,10,10bis}. The confinement of gauge
fields on lower-dimensional hypersurfaces was also investigated \cite
{11,11ter}. More recent models even suggest that all Standard Model
particles could be confined on the branes \cite{11bis}. Therefore, finding
physical evidences of extra dimensions is a major contemporary challenge.
Interesting results could arise from high energy physics (Kaluza-Klein tower
states \cite{12} for instance) or low energy physics (deviations from the
inverse square law of gravity \cite{13} for instance).

In the present paper, we are mainly motivated by the quest of new phenomena
at a non-relativistic energy scale. We explore how the quantum dynamics of
fermions is modified when the higher-dimensional bulk contains more than
only one brane. Hence, we focus on a two-brane world (related to two
domain-walls) and investigate the dynamics of a massive fermion in this
extended framework. It is shown that such a model is formally equivalent to
a two-sheeted spacetime (a product manifold $M_4\times Z_2$) described in
the formalism of the noncommutative geometry \cite{14,15,16,17}, at least as
a low-energy effective theory.

In previous works \cite{23,24}, the present authors have studied the
phenomenology of certain of these $M_4\times Z_2$ two-sheeted spacetimes,
but no formal proof had been given that these exotic geometries could be
related to more conventional branes theories: the link between both
approaches was just considered as a fairly working hypothesis. For the first
time, in the present paper, a physical and mathematical proof of this link
is derived. Moreover, the mathematical description of the $M_4\times Z_2$
geometry is enlarged by contrast to the previous works \cite{23,24}. The
demonstration made in the present paper is inspired by quantum chemistry and
the construction of molecular orbitals, here extended to branes. As a
consequence of the bulk dimensionality extension, the quantum dynamics
phenomenology is considerably enriched: for a 5D bulk containing two branes,
matter swapping between these two worlds is made possible (although the
effect could remain difficult to observe). More important, since the
obtained equations (extended Dirac and Pauli's equations) are completely
model-independent, we conclude that this matter swapping phenomenon might
probably be shared by every model of universe containing at least two
''worlds''.

In section II, we introduce the kink-antikink domain-walls
description of a two-brane world. Section III gives the fermion
eigenstates in such a braneworld setup. In section IV, we
introduce gauge fields in the two-brane world model. In section V,
we then derive a two-level description of the fermion dynamics in
a braneworld with two domain-walls in presence of an
electromagnetic field. In section VI, we show that the two-level
description fits with that of a two-sheeted spacetime as described
by noncommutative geometry. Finally, in section VII we underline
the basic phenomenological consequences of the present work.

\section{Braneworlds with two domain-walls}

The brane concept takes its origin in superstring theories \cite{18}, though
earlier similar concepts were proposed in other theoretical contexts \cite
{19,20}. However, since superstring theories suffer from mathematical
complications, several simplified approaches relying on more classical field
theories \cite{21} have been suggested. For instance, several works \cite
{5,6,7,8,9,10,10bis} are now inspired by the approach of Rubakov and
Shaposhnikov \cite{20}. These authors have suggested that elementary
particles might be trapped on a defect (a domain-wall) on a
higher-dimensional spacetime. Such a defect would arise from a scalar field
whose dynamics should be described by a soliton-kink solution in a $\varphi
^4$ theory. Bosonic excitations of the scalar field are trapped and
propagate along the kink. In addition, any chiral Dirac particle is also
trapped on the domain-wall \cite{5,6,20}. More recently, Randall and Sundrum
have suggested braneworlds models where the bulk metric is warped to ensure $%
(3+1)$-dimensional gravity to be reproduced as well \cite{22}.

The model considered in the present paper relies on a usual brane-world
description involving domain-walls in the bulk \cite{5,6}. Two branes are
here considered living in a continuous 5D manifold. Branes are described by
the kink (antikink) solutions of a scalar field in a $\varphi ^4$ theory.
Matter is then described through a 5D fermionic field coupled to this scalar
field. Since we are not motivated by gravitational considerations, we use a
flat metric for the bulk spacetime, i.e.:

\begin{equation}
ds^2=g_{AB}dx^Adx^B=g_{\mu \nu }dx^\mu dx^\nu -dz^2=\eta _{\mu \nu }dx^\mu
dx^\nu -dz^2  \label{1}
\end{equation}
where $g_{AB}$ is the five-dimensional metric tensor with signature $%
(+,-,-,-,-)$ with $A,B=0,\ldots ,4$. $\eta _{\mu \nu }$ is the
four-dimensional Minkowski metric tensor of signature $(+,-,-,-)$ with $\mu
,\nu =0,\ldots ,3$, and $z$ the coordinate along the extra dimension. An
improved model involving a warped metric could be considered as well \cite
{5,6,7}, but this choice would introduce supplementary complications which
are neither relevant nor necessary to illustrate the mechanism discussed in
this paper. The action $S$ for a real scalar field $\Phi $ coupled to a
matter field $\Psi $ in a five-dimensional spacetime is then
\begin{equation}
S=\int \left[ \frac 12g^{AB}\left( \partial _A\Phi \right) \left( \partial
_B\Phi \right) -V(\Phi )+\overline{\Psi }\left( i\Gamma ^A\partial
_A-\lambda \Phi \right) \Psi \right] \sqrt{g}d^5x  \label{2}
\end{equation}
We assume that the Dirac field $\Psi $ is coupled to the scalar field $\Phi $
through a Yukawa coupling term $\lambda \overline{\Psi }\Phi \Psi $ where $%
\lambda $ is the coupling constant. It should be pointed out that another
choice for the coupling term \cite{8} would not change the final conclusions
of the paper. In addition, a convenient potential $V(\Phi )$ is given by
\cite{5,6,20}:
\begin{equation}
V(\Phi )=\frac \chi 4(\Phi ^2-\eta ^2)^2  \label{3}
\end{equation}
where $\chi $ and $\eta $ are constants of the potential. Though several
possibilities can be considered for the potential \cite{5,6,8}, we just
assume that it allows the existence of domain-walls (i.e. topological
defects) in accordance with the original Rubakov-Shaposhnikov concept \cite
{20}. The scalar field equations of motion can be easily derived from
relation (\ref{2}):

\begin{equation}
\Phi ^{\prime \prime }+\chi \eta ^2\Phi -\chi \Phi ^3=0  \label{4}
\end{equation}
from which domain-wall solutions can then be derived. $\Phi ^{\prime \prime
} $ is the second order derivative of $\Phi$ along the extra dimension.

For a single brane, two solutions have to be considered:
\begin{equation}
\Phi _{\frac k{ak}}(z)=\pm \Phi (z)=\pm \eta \tanh \left( z/\xi \right)
\label{5}
\end{equation}
where ''$+$'' (respectively ''$-$'') refers to the kink ($k$) solution
(respectively antikink ($ak$) solution). $\xi $ is the brane thickness such
that $\xi ^{-1}=\eta \sqrt{\chi /2}$.

Now if we shift from a single brane to a two-brane world, the solution can
be expressed as a kink-antikink pair, each wall being localized respectively
at $z=-d/2$ and $z=+d/2$. The field solution of eq. (\ref{4}), which
describes such a kink-antikink system, can be approximated by \cite{10,10bis}
\begin{eqnarray}
\Phi (z) &=&\Phi _{-}(z)+\Phi _{+}(z)+\Delta \Phi  \label{6} \\
&=&\Phi (z+d/2)-\Phi (z-d/2)-\eta  \nonumber
\end{eqnarray}
provided that $d\gg \xi $ i.e. the distance $d$ between branes is larger
than the brane thickness. For that reason, we are now considering that both
branes are independent from each other both from a scalar and a
gravitational point of view.

\section{Fermions in braneworlds}

From eq. (\ref{2}), the five-dimensional massless Dirac equation can be
easily expressed. In what follows, the Dirac matrices are given by $\Gamma
^\mu =\gamma ^\mu $ and $\Gamma ^4=-i\gamma ^5=\gamma ^0\gamma ^1\gamma
^2\gamma ^3$, where $\gamma ^\mu $ and $\gamma ^5$ are the usual Dirac
matrices in the four-dimensional Minkowski spacetime. The Clifford algebra
is verified since
\begin{equation}
\left\{ \Gamma ^A,\Gamma ^B\right\} =2\eta ^{AB}  \label{7}
\end{equation}
where $\eta ^{AB}$ is the five-dimensional metric tensor of the Minkowski
spacetime. The Dirac equation is therefore:
\begin{equation}
\left( i\gamma ^\mu \partial _\mu +\gamma ^5\partial _z-\lambda \Phi \right)
\Psi =0  \label{8}
\end{equation}
By using the separating variable method and due to $\gamma ^5$ matrix in eq.
(\ref{8}) the solution $\Psi $ can be expressed as
\begin{equation}
\Psi (x,z)=f_L(z)\psi _L(x)+f_R(z)\psi _R(x)  \label{9}
\end{equation}
where the $\psi _{L,R}$ are left- and right-handed 4D spinors such that $%
\gamma ^5\psi _{R/L}=\pm \psi _{R/L}$. $x$ are the four-dimensional
coordinates. For a trapped fermion, we expect that the five-dimensional
Dirac equation can be expressed as an effective four-dimensional massive
equation such that:
\begin{equation}
i\gamma ^\mu \partial _\mu \psi _{L/R}=m\psi _{R/L}  \label{10}
\end{equation}
where $m$ is assumed to be the apparent (four-dimensional) particle mass.
Substituting eq. (\ref{9}) into eq. (\ref{8}), we get then
\begin{equation}
\left\{
\begin{array}{c}
\partial _zf_R-\lambda \Phi f_R+mf_L=0 \\
\partial _zf_L+\lambda \Phi f_L-mf_R=0
\end{array}
\right.  \label{11}
\end{equation}
After a convenient rearrangement of the equations, we get:
\begin{equation}
\left\{
\begin{array}{c}
-\partial _{zz}f_L(z)+W_Lf_L(z)=m^2f_L(z) \\
-\partial _{zz}f_R(z)+W_Rf_R(z)=m^2f_R(z)
\end{array}
\right.  \label{12}
\end{equation}
with
\begin{equation}
\left\{
\begin{array}{c}
W_L=\lambda \left( \lambda \Phi ^2-\left( \partial _z\Phi \right) \right) \\
W_R=\lambda \left( \lambda \Phi ^2+\left( \partial _z\Phi \right) \right)
\end{array}
\right.  \label{13}
\end{equation}
and
\begin{equation}
\int f_\alpha (z)f_\beta (z)dz=\delta _{\alpha ,\beta }\text{ and }\int
\left( \psi _L^{\dagger }(x)\psi _L(x)+\psi _R^{\dagger }(x)\psi
_R(x)\right) d^4x=1  \label{14}
\end{equation}
Obviously, owing to the Schrodinger-like equations (\ref{12}), $f_{L,R}(z)$
define the localization of the left- and right-handed states of the fermion
along the extra dimension with effective potentials $W_{L/R}$. $m$ is the
effective mass of the trapped fermion and it is related to the eigenvalues
of the bound states in the potentials $W_{L/R}$. Eqs. (\ref{13}) imply that
the effective potential felt by the fermion depends on its helicity state.
Then, left- and right-handed states are not necessarily localized at the
same place and it is even possible that bound state cannot exist. Previous
works \cite{5,6,20} have shown that for a kink-solution of the scalar field,
the $m=0$ mode is localized on the kink for left-handed state whereas
right-handed state cannot be localized. By contrast, the $m\neq 0$ modes are
localized around the kink whatever their state. In an antikink world, the $%
m\neq 0$ modes are also localized around the antikink whatever their state,
but in the opposite only right-handed $m=0$ fermions can exist. Note that
since the usual $4D$ fermion wave-function $\psi (x)$ can be expressed as
\begin{equation}
\psi (x)=\psi _L(x)+\psi _R(x)  \label{15}
\end{equation}
it can then be easily shown that eq. (\ref{9}) can be rewritten as
\begin{equation}
\Psi (x,z)=\Pi (z)\psi (x)  \label{16}
\end{equation}
where
\begin{eqnarray}
\Pi (z)=f(z)+\gamma ^5\kappa (z)  \label{17}
\end{eqnarray}
with
\begin{equation}
\left\{
\begin{array}{c}
\kappa (z)=\left( 1/2\right) \left( f_R(z)-f_L(z)\right) \\
f(z)=\left( 1/2\right) \left( f_R(z)+f_L(z)\right)
\end{array}
\right.  \label{18}
\end{equation}
and
\begin{equation}
\int \Pi ^{\dagger }(z)\Pi (z)dz=\mathbf{1}_{4\times 4}  \label{19}
\end{equation}
where $\Pi (z)$ defines the localization of the fermion along the extra
dimension for any helicity state. We note that $\overline{\Psi }(x,z)=%
\overline{\psi }(x)\overline{\Pi }(z)$ with $\overline{\Pi }(z)=f(z)-\gamma
^5\kappa (z)$.

\subsection{Fermionic wave functions in a single brane-world}

\label{sec3A}

A single brane-world solution for a trapped fermion can be easily derived
from eqs. (\ref{12}). By first considering a single kink domain-wall, the
effective potential derived from eqs. (\ref{13}) and eq. (\ref{5}) becomes:
\begin{equation}
W_{L/R}^{1 kB}(z)=\frac 1{\xi ^2}\left\{ \varepsilon ^2-\varepsilon \left(
\varepsilon \pm 1\right) \frac 1{\cosh ^2\left( z/\xi \right) }\right\}
\label{20}
\end{equation}
where $\varepsilon =\lambda \sqrt{2/\chi }=\lambda \eta \xi $. Expression (%
\ref{20}) corresponds to a P\"{o}schl-Teller potential for which eqs. (\ref
{12}) present well-known analytical solutions \cite{6}. Let us recall the
first two modes:

$\bullet $ For $m=m_0=0$%
\begin{equation}
\left\{
\begin{array}{c}
f_{0,L}=N_0\cosh ^{-\varepsilon }\left( z/\xi \right) \\
f_{0,R}=0
\end{array}
\right.  \label{21}
\end{equation}

$\bullet $ For $m=m_1=(1/\xi )\sqrt{2\varepsilon -1}$%
\begin{equation}
\left\{
\begin{array}{c}
f_{1,L}=N_1\cosh ^{-\varepsilon }\left( z/\xi \right) \sinh \left( z/\xi
\right) \\
f_{1,R}=N_2\cosh ^{-\varepsilon +1}\left( z/\xi \right)
\end{array}
\right.  \label{22}
\end{equation}
with
\begin{eqnarray}
N_0=\sqrt{\frac{\Gamma (\varepsilon +1/2)}{\xi \sqrt{\pi }\Gamma
(\varepsilon )}}\text{, }N_1=\sqrt{2\varepsilon -2}\sqrt{\frac{\Gamma
(\varepsilon +1/2)}{\xi \sqrt{\pi }\Gamma (\varepsilon )}}\text{, and }N_2=%
\sqrt{\frac{\Gamma (\varepsilon -1/2)}{\xi \sqrt{\pi }\Gamma (\varepsilon -1)%
}}  \label{23}
\end{eqnarray}
where $\Gamma (x)$ is the usual Gamma function. Obviously, $\varepsilon $
behaves like a coupling constant between the brane and the fermion. The
trapping mechanism becomes more and more effective when $\varepsilon $
increases (the spatial extensions of solutions (\ref{21}) and (\ref{22})
decrease when $\varepsilon $ increases). For an antikink-brane, the
solutions can be easily deduced from the previous ones through a simple $%
L\leftrightarrow R$ substitution.

\subsection{Fermionic wave functions in a two-brane world}

A system of two branes can be described by a two-well effective potential $%
W_{L,R}^{2B}$ derived from eqs. (\ref{6}) and (\ref{13}). The condition $%
d\gg \xi $ implies that the distance between the two wells is large. When $%
d\rightarrow +\infty $, each well becomes a local one and it behaves as if
there was a single kink (or antikink) in the bulk. In that case, we obtain
the local potentials $W_{L/R}^{1kB}(z+d/2)$ (or $W_{L/R}^{1akB}(z-d/2)$)
resulting from a single kink (or antikink) distant from the antikink (or
kink). If we assume that each brane should possess its own copy of the
standard model, it is then legitimate to build the two-brane fermionic
solutions from the local one-brane fermionic solutions. This way to proceed
is similar to atomic orbital combination used in quantum chemistry to build
molecular orbitals. Similarly, for a system of two branes we define the
global fermion state as:
\begin{equation}
\Psi (x,z)=\Pi _{+}(z)\psi _{+}(x)+\Pi _{-}(z)\psi _{-}(x)  \label{24}
\end{equation}
where $\pm $ denote $z=\pm d/2$ i.e. the location of each brane. The states $%
\Pi _{\pm }(z)$ correspond to the fermion eigenstates related to the branes $%
(+)$ and $(-)$ considered here as independent from each other. Obviously, $%
\psi _{\pm }(x)$ will differ from the solutions $\psi (x)$ obtained by
solving eq. (\ref{10}) for a single brane-world. The states $\Pi _{\pm }(z)$
can be easily deduced from the one-brane fermionic solutions (see subsection
\ref{sec3A}). $\Pi _{-}(z)$ will use kink-brane fermionic solutions with a
translation $z\rightarrow z+d/2$ while $\Pi _{+}(z)$ will use antikink-brane
fermionic solutions with a translation $z\rightarrow z-d/2$.

\section{Gauge fields in domain-walls}

\label{GF}

In the following, we consider the introduction of gauge fields in the model
with a special emphasis on electromagnetism. Localizing gauge fields on a
domain-wall remains a delicate task \cite{11,11ter,11bis}. Among the
numerous approaches proposed to localize gauge fields on branes \cite
{11,11ter,11bis}, the approach of Dvali, Gabadadze, Porrati and Shifman
(DGPS) \cite{11ter} is quite generic and model-independent. The basic
ingredient is a bulk vectorial field (a photon like field) which is
minimally coupled to some of the matter fields localized on a brane. Thus,
introducing a $U(1)$ gauge field $\mathcal{A}$ in eq. (\ref{2}) the
five-dimensional action becomes then:
\begin{equation}
S=\int \left[ -\frac 1{4G^2}\mathcal{F}_{AB}\mathcal{F}^{AB}+\frac
12g^{AB}\left( \partial _A\Phi \right) \left( \partial _B\Phi \right)
-V(\Phi )+\overline{\Psi }\left( i\Gamma ^A\left( \partial _A+i\mathcal{A}%
_A\right) -\lambda \Phi \right) \Psi \right] \sqrt{g}d^5x  \label{4.1}
\end{equation}
where $G$ is a coupling constant.

Through the quantum fluctuations of the five-dimensional gauge
field, the localized fermionic fields induce gauge field
localization. Indeed, the gauge field propagator receives
corrections from one-loop diagrams with localized matter fields
running in the loops. This leads to a four-dimensional kinetic
term which results from the need of a counter-term in the
five-dimensional gauge field Lagrangian. An effective
four-dimensional gauge field theory results on the brane. The bulk
field is then forced to propagate along the three-dimensional
space of the brane, at
least for distances lower than a critical cosmological distance \cite{11ter}%
. The same procedure can be used with more complex domain-wall approaches
(including those relying on warped metric) where other phenomena can
contribute to gauge fields confinement on branes \cite{11ter}.

\subsection{Gauge field in a single brane-world}

\label{OBGF}

From eq. (\ref{4.1}) the 5D interaction action between the matter
field and the $U(1)$ gauge field takes the form:
\begin{eqnarray}
S_{\text{int}}=-\int d^4xdzJ_A(x,z)\mathcal{A}^A(x,z)=-\int d^4xdz%
\overline{\Psi }(x,z)\Gamma _A\Psi (x,z)\mathcal{A}^A(x,z)  \label{4.2}
\end{eqnarray}
The four-dimensional current is $j^\mu (x)=\overline{\psi }(x)\gamma ^\mu
\psi (x)$, with $j^\mu (x)=\int J^\mu (x,z)dz$ and $\int J^5(x,z)dz=0$.
Using eqs. (\ref{16}) to (\ref{18}), it can be shown that $\int \partial
_AJ^A(x,z)dz=\partial _\mu j^\mu (x)$. From the five-dimensional current
conservation $\partial _AJ^A(x,z)=0$, as $\int \partial _AJ^A(x,z)dz=0$, we
deduce then that $\partial _\mu j^\mu (x)=0$, i.e. the four-dimensional
current is conserved. This implies the transversality of the loop. Moreover,
we chose the Lorentz gauge in the bulk
\begin{equation}
\partial _A\mathcal{A}^A(x,z)=0  \label{4.3}
\end{equation}
From the four-dimensional transversality of currents we get $\partial _\mu
\mathcal{A}^\mu (x,z)=0$ \cite{11ter} and then from eq. (\ref{4.3}) we
deduce that $\partial _z\mathcal{A}^z(x,z)=0$, i.e. $\left. \mathcal{A}%
^z(x,z)\right| _{x=Cte}=Cte$. As a consequence, the remaining
relevant interaction action between the localized matter field and
the bulk vector field is then:
\begin{eqnarray}
S_{\text{int}}=-\int d^4x\overline{\psi }(x)\gamma _\mu \psi
(x)a^\mu (x)  \label{4.4}
\end{eqnarray}
where an effective four-dimensional vector field $a^\mu (x)$ can be defined
as
\begin{equation}
a^\mu (x)=(1/2)\int \left\{ f_R^2(z)+f_L^2(z)\right\} \mathcal{A}^\mu (x,z)dz
\label{4.5}
\end{equation}
$a^\mu (x)$ acts as a $U(1)$ gauge field in a four-dimensional spacetime.
The interaction Lagrangian leads to a supplementary kinetic term induced by
localized fermionic one-loop diagrams with two external $a_\mu (x)$ legs
\cite{11ter}. The low-energy action on the brane must then contain the
induced term:
\begin{equation}
-\frac 1{4e^2}F_{\mu \nu }F^{\mu \nu }  \label{4.6}
\end{equation}
with
\begin{equation}
F_{\mu \nu }=\partial _\mu a_\nu -\partial _\nu a_\mu   \label{4.7}
\end{equation}
where $e$ is the effective coupling constant. Rigorously other corrective
terms should be also considered in the action, but we do not discuss them
here. Moreover, we are not considering the details of the propagation of the
gauge field in the bulk or onto the brane (see references \cite{11ter}). The
matters have already been considered in details in previous works \cite
{11ter}. We just note that the separating variable method leads to write
\begin{equation}
\mathcal{A}(x,z)=\Lambda (z)\mathcal{A}(x)  \label{4.8}
\end{equation}
with $\Lambda (z)$ a function that quickly decreases when moving away from
the branes \cite{11ter}.

\subsection{Gauge field in a two-brane world}

\label{TBGF}

Let us now consider the introduction of gauge fields in our two-brane world.
As previously explained for fermions, we assume that each brane possesses
its own copy of the standard model. The two-brane gauge field solutions can
be derived from the local one-brane gauge solutions such that:

\begin{eqnarray}
\mathcal{A}(x,z) &=&\mathcal{A}^{+}(x,z)+\mathcal{A}^{-}(x,z)  \label{4.9} \\
&=&\Lambda _{+}(z)\mathcal{A}_{+}(x)+\Lambda _{-}(z)\mathcal{A}_{-}(x)
\nonumber
\end{eqnarray}
where $\pm $ denote $z=\pm d/2$ i.e. the location of each brane. The states $%
\Lambda _{\pm }(z)$ correspond to the gauge field localized states related
to the branes $(+)$ and $(-)$ considered here as independent from each
other. Obviously, $\mathcal{A}_{\pm }(x)$ will differ from the solutions $%
\mathcal{A}(x)$ in eq. (\ref{4.8}) for a single brane-world. The states $%
\Lambda _{\pm }(z)$ can be easily deduced from the one-brane gauge field
solutions. $\Lambda _{-}(z)$ ($\Lambda _{+}(z)$) is related to kink-brane
(antikink-brane) gauge field solutions with a translation $z\rightarrow
z+d/2 $ ($z\rightarrow z-d/2$).

On each brane, we get the local four-dimensional vector field:
\begin{equation}
a_\mu ^{\pm }(x)=(1/2)\int \left\{ f_{R,\pm }^2(z)+f_{L,\pm }^2(z)\right\}
\mathcal{A}_\mu ^{\pm }(x,z)dz  \label{4.9bis}
\end{equation}

From eq. (\ref{4.2}), it must be noted that the two-brane description of the
gauge field and of the fermionic field leads to specific cross-terms. For
instance, we get
\begin{equation}
(1/2)\int \left\{ f_{R,+}^2(z)+f_{L,+}^2(z)\right\} \mathcal{A}_\mu
^{-}(x,z)dz  \label{4.10}
\end{equation}
which can be interpreted as the four-dimensional gauge field induced in the
brane $(+)$ by charges localized in the brane $(-)$. In fact, a simple
analysis shows that this term is proportional to $\exp (-2\left( \varepsilon
-1\right) d/\xi )$, i.e. a charge localized in a brane acts as a kind of
''millicharged'' particle in the second brane. For instance, with $%
\varepsilon =2$ and $d/\xi =22$ (see discussion in appendix \ref{appendixB}%
), a charge $q_e$ localized in the other brane would act in our brane as an
effective particle with a charge $q=10^{-19}q_e$. For such tiny values, the
effect can be neglected \cite{25a}.

\section{Two-level approximation of fermion dynamics in a braneworld with
two domain-walls}

\label{sec4}

Let us now show that the above two-brane world model reduces to a simple
two-level quantum description. At low energy, a brane can be assumed to be
an infinitely thin 4D sheet where SM particles live. Therefore, for a single
kink (antikink)-brane, the projection of eq. (\ref{8}) onto its $f_{L,R}$
eigenstates, will reduce the 5D Dirac equation to a 4D equation with a mass $%
m$ particle located at $z=-d/2$ for instance (or $z=d/2$). The projection is
equivalent to a dimensional reduction leading to a single 4D Dirac equation.
Similarly, for a system of two thin branes, the projection of eq. (\ref{8})
on the eigenstates of each independent brane, will lead to two coupled 4D
Dirac equations. Although this approach is quite unusual in the present
context, it is perfectly well founded. It is exactly the procedure used in
quantum chemistry to approximate molecular orbitals by solving the
Hamiltonian in the subspace of each atomic eigenstates. Here, the
Hamiltonian for a two brane-world is expressed by using the fermionic
eigenstates of each independent branes. This approximation is valid as long
as both branes are distant enough in the bulk. Let us now derive the
resulting system of 4D coupled Dirac equations.

Taking account of the electromagnetic gauge vector field (see eq. (\ref{4.1}%
)), the five-dimensional Dirac equation can be expressed in a Schrodinger
form:
\begin{equation}
i\partial _0\Psi =H\Psi  \label{25}
\end{equation}
with
\begin{equation}
H=-i\gamma ^0\gamma ^\eta \left( \partial _\eta +i\mathcal{A}_\eta \right)
-\gamma ^0\gamma ^5\left( \partial _z+i\mathcal{A}_z\right) +\gamma
^0\lambda \Phi +\mathcal{A}_0  \label{26}
\end{equation}
where $\eta =1,2,3$. In the following, we will consider a specific mass
state and we assume that there is no mixing, coupling or interaction between
this state and other fermion states of different mass. Therefore, the states
$\psi _{+}(x)$ and $\psi _{-}(x)$ exhibit the same mass. Moreover, since the
terms of higher mass are neglected, our approach remains clearly an
approximation of low energy. Using the expression of $\Psi $ given in eq. (%
\ref{24}), eq. (\ref{25}) is projected onto the localized states $\Pi _{\pm
}(z)$. Introducing then:
\begin{equation}
\left\{
\begin{array}{c}
h_{i,j}=\int \Pi _i^{\dagger }(z)H\Pi _j(z)dz \\
s=s^{\dagger }=\int \Pi _{+}^{\dagger }(z)\Pi _{-}(z)dz
\end{array}
\right.  \label{27}
\end{equation}
and using a convenient matrix representation, one obtains easily:
\begin{equation}
i\partial _0\left(
\begin{array}{cc}
1 & s \\
s & 1
\end{array}
\right) \left(
\begin{array}{c}
\psi _{+}(x) \\
\psi _{-}(x)
\end{array}
\right) =\left(
\begin{array}{cc}
h_{+,+} & h_{+,-} \\
h_{-,+} & h_{-,-}
\end{array}
\right) \left(
\begin{array}{c}
\psi _{+}(x) \\
\psi _{-}(x)
\end{array}
\right)  \label{28}
\end{equation}
Considering
\begin{equation}
\Psi =\left(
\begin{array}{c}
\psi _{+}(x) \\
\psi _{-}(x)
\end{array}
\right)  \label{29}
\end{equation}
the two-level Dirac-Schrodinger equation
\begin{equation}
i\partial _0\Psi =\widetilde{H}\Psi  \label{30}
\end{equation}
can be easily deduced from eq. (\ref{28}) with the two-level Hamiltonian $%
\widetilde{H}$ given by:
\begin{equation}
\widetilde{H}=\frac 1{1-s^2}\otimes \left(
\begin{array}{cc}
h_{+,+}-sh_{-,+} & h_{+,-}-sh_{-,-} \\
h_{-,+}-sh_{++} & h_{-,-}-sh_{+,-}
\end{array}
\right)  \label{31}
\end{equation}
Using equations (\ref{10}), (\ref{11}), (\ref{14})-(\ref{18}), (\ref{26})
and (\ref{27}), the terms in eq. (\ref{31}) can be simplified (see appendix
\ref{appendixA}) to give:
\begin{equation}
\widetilde{H}=\left(
\begin{array}{cc}
-i\gamma ^0\gamma ^\eta (\partial _\eta +iqA_\eta ^{+})+\gamma ^0m+\gamma
^0\delta m+qA_0^{+} & -\gamma ^0\gamma ^5g+\gamma ^0m_r-\gamma ^0\gamma
^5\Upsilon \\
\gamma ^0\gamma ^5g+\gamma ^0m_r+\gamma ^0\gamma ^5\overline{\Upsilon } &
-i\gamma ^0\gamma ^\eta (\partial _\eta +iqA_\eta ^{-})+\gamma ^0m+\gamma
^0\delta m+qA_0^{-}
\end{array}
\right)  \label{32}
\end{equation}
with $A_\mu ^{\pm }$ the electromagnetic fields of the brane $(+)$ or $(-)$,
$q$ the electric charge of the fermion, and where
\begin{eqnarray}
\left\{
\begin{array}{c}
g=\lambda \int \left\{ \Phi _{-}+\Delta \Phi \right\} \left\{ f_{-}(z)\kappa
_{+}(z)-\kappa _{-}(z)f_{+}(z)\right\} dz \\
m_r=\lambda \int \left\{ \Phi _{-}+\Delta \Phi \right\} \left\{
f_{-}(z)f_{+}(z)-\kappa _{-}(z)\kappa _{+}(z)\right\} dz \\
\delta m=\lambda \int \Phi _{-}\left\{ f_{+}^2(z)-\kappa _{+}^2(z)\right\} dz
\end{array}
\right.  \label{33}
\end{eqnarray}
and
\begin{equation}
\left\{
\begin{array}{c}
\Upsilon =\varphi +\gamma ^5\phi \\
\overline{\Upsilon }=\varphi ^{*}-\gamma ^5\phi ^{*}
\end{array}
\right.  \label{35bis}
\end{equation}
where $\varphi $ and $\phi $ are the scalar components of the off-diagonal
part $\Upsilon $ of the effective gauge field such that
\begin{equation}
\left\{
\begin{array}{c}
\phi =i\int \left\{ f_{+}(z)\kappa _{-}(z)-\kappa _{+}(z)f_{-}(z)\right\}
\left\{ \mathcal{A}_z^{+}(x,z)+\mathcal{A}_z^{-}(x,z)\right\} dz \\
\varphi =i\int \left\{ f_{+}(z)f_{-}(z)-\kappa _{+}(z)\kappa _{-}(z)\right\}
\left\{ \mathcal{A}_z^{+}(x,z)+\mathcal{A}_z^{-}(x,z)\right\} dz
\end{array}
\right.  \label{35ter}
\end{equation}
An interpretation of those off-diagonal gauge terms as well as the way to
deal with them will be discussed in the next section.

Let us apply a convenient $SU(2)$ rotation such that
\begin{equation}
\left(
\begin{array}{c}
\psi _{+}(x) \\
\psi _{-}(x)
\end{array}
\right) \rightarrow \left(
\begin{array}{cc}
e^{-i\pi /4} & 0 \\
0 & e^{i\pi /4}
\end{array}
\right) \left(
\begin{array}{c}
\psi _{+}(x) \\
\psi _{-}(x)
\end{array}
\right)  \label{34}
\end{equation}
Back to the Dirac form, equation (\ref{30}) then reads:
\begin{equation}
\left(
\begin{array}{cc}
i\gamma ^\mu (\partial _\mu +iqA_\mu ^{+})-m-\delta m & ig\gamma
^5-im_r+i\gamma ^5\Upsilon \\
ig\gamma ^5+im_r+i\gamma ^5\overline{\Upsilon } & i\gamma ^\mu (\partial
_\mu +iqA_\mu ^{-})-m-\delta m
\end{array}
\right) \Psi =0  \label{35}
\end{equation}

As a consequence of the first order approximation (see appendix \ref
{appendixA}), $g$ is only related to the first order derivative $\partial _z$
along the continuous extra dimension while $m_r$ is only related to the
scalar field $\Phi $. From solutions of eq. (\ref{11}) (see subsection \ref
{sec3A}), $g$, $m_r$ and $\delta m$ can be easily estimated as shown in
appendix \ref{appendixB}.

Let us now show that eq. (\ref{35}) describing the dynamics of a fermion in
a two domain-walls setup is equivalent to that of a particle embedded in a
noncommutative two-sheeted spacetime.

\section{''Noncommutative'' two-sheeted spacetime interpretation of the
two-level approximation}

\subsection{Noncommutative two-sheeted spacetime}

In refs. \cite{23,24}, a model describing the quantum dynamics of fermions
in a two-sheeted spacetime (i.e. a two-brane world) has been proposed. Such
a universe corresponds to the product of a four-dimensional continuous
manifold with a discrete two-point space and can be seen as a 5D universe
with a fifth dimension reduced to two points with coordinates $\pm \delta /2$
(both sheets are separated by a phenomenological distance $\delta $).
Mathematically, the model relies on a bi-euclidean space $X=M_4\times Z_2$
in which any smooth function belong to the algebra $A=C^\infty (M)\oplus
C^\infty (M)$ and can be adequately represented by a $2\times 2$ diagonal
matrix $F$ such that:
\begin{equation}
F=\left(
\begin{array}{cc}
f_1 & 0 \\
0 & f_2
\end{array}
\right)  \label{36}
\end{equation}
In the noncommutative formalism, the expression of the exterior derivative $%
D=d+Q$, where $d$ acts on $M_4$ and $Q$ on the $Z_2$ internal variable, has
been given by A. Connes \cite{14}: $D:(f_1,f_2)\rightarrow
(df_1,df_2,g(f_2-f_1),g(f_1-f_2))$ with $g=1/\delta $. Viet and Wali \cite
{16} have proposed a representation of $D$ acting as a derivative operator
and fulfilling the above requirements (see also \cite{17}). Due to the
specific geometrical structure of the bulk, this operator is given by:

\begin{equation}
D_\mu =\left(
\begin{array}{cc}
\partial _\mu & 0 \\
0 & \partial _\mu
\end{array}
\right) ,\text{ }\mu =0,1,2,3\text{ and\ }D_5=\left(
\begin{array}{cc}
0 & g \\
-g & 0
\end{array}
\right)  \label{37}
\end{equation}
Where the term $g$ acts as a finite difference operator along the discrete
dimension. Using (\ref{37}), one can build the Dirac operator defined as
\begin{equation}
\not{D}=\Gamma ^ND_N=\Gamma ^\mu D_\mu +\Gamma ^5D_5  \label{38}
\end{equation}
By considering the following extension of the gamma matrices (we are working
in the Hilbert space of spinors, see \cite{14})
\begin{equation}
\Gamma ^\mu =\left(
\begin{array}{cc}
\gamma ^\mu & 0 \\
0 & \gamma ^\mu
\end{array}
\right) \text{\ and\ }\Gamma ^5=\left(
\begin{array}{cc}
\gamma ^5 & 0 \\
0 & -\gamma ^5
\end{array}
\right)  \label{39}
\end{equation}
where $\gamma ^\mu $ and $\gamma ^5=i\gamma ^0\gamma ^1\gamma ^2\gamma ^3$
are the usual Dirac matrices, it can be easily shown that the Dirac operator
given by eq. (\ref{38}) has the following self adjoint realization:
\begin{equation}
\not{D}=\left(
\begin{array}{cc}
{\not{D}}_{+} & g\gamma ^5 \\
g\gamma ^5 & {\not{D}}_{-}
\end{array}
\right) =\left(
\begin{array}{cc}
\gamma ^\mu \partial _\mu & g\gamma ^5 \\
g\gamma ^5 & \gamma ^\mu \partial _\mu
\end{array}
\right)  \label{40}
\end{equation}
In some ''noncommutative'' two-sheeted models \cite{15}, the off-diagonal
terms proportional to $g$ are often related to the particle mass through the
Higgs field. As shown in previous works \cite{23}, $g$ can also be
considered as a constant geometrical field and takes the same value for each
particle. We can therefore introduce a mass term as in the standard Dirac
equation, but more general, i.e.:

\begin{equation}
M=\left(
\begin{array}{cc}
m\mathbf{1}_{4\times 4} & m_c\mathbf{1}_{4\times 4} \\
m_c^{*}\mathbf{1}_{4\times 4} & m\mathbf{1}_{4\times 4}
\end{array}
\right)  \label{41}
\end{equation}
where ''$*$'' denotes the complex conjugate. The two-sheeted Dirac equation
writes:
\begin{eqnarray}
{\not{D}}_{dirac}\Psi &=&\left( {i\not{D}-M}\right) \Psi =\left( {i\Gamma
^ND_N-M}\right) \Psi =  \label{42} \\
&=&\left(
\begin{array}{cc}
i\gamma ^\mu \partial _\mu -m & ig\gamma ^5-m_c \\
ig\gamma ^5-m_c^{*} & i\gamma ^\mu \partial _\mu -m
\end{array}
\right) \left(
\begin{array}{c}
\psi _{+} \\
\psi _{-}
\end{array}
\right) =0  \nonumber
\end{eqnarray}
with $\Psi =\left(
\begin{array}{c}
\psi _{+} \\
\psi _{-}
\end{array}
\right) $ the two-sheeted wave function. In this notation, the indices ``$+$%
'' and ``$-$'' are purely conventional and simply allow to discriminate the
two sheets (or branes) embedded in the 5D bulk. It can be noticed that by
virtue of the two-sheeted structure of spacetime, the wave function $\psi $
of the fermion is split into two components, each component living on a
distinct spacetime sheet. If one considers the $(-)$ sheet to be our brane,
the $\psi _{+}$ part of the wave function, in the $(+)$ sheet, can be
considered as a hidden particle component.

\subsection{Gauge fields in the two-sheeted spacetime}

Pursuing with the approach introduced in the section \ref{GF}, we are now
illustrating how the electromagnetic fields behaves in the present
formalism. It should be emphasized that the results presented here for
electromagnetism could be extended to other interactions as well, especially
electroweak interactions and chromodynamics. To be consistent with the
structure of the Dirac field $\Psi $ in eq. (\ref{42}), the usual $U(1)$
electromagnetic gauge field has to be replaced by an extended $U(1)\otimes
U(1)$ gauge field. The group representation is therefore:
\begin{equation}
G=\left(
\begin{array}{cc}
\exp (-iq\Lambda _{+}) & 0 \\
0 & \exp (-iq\Lambda _{-})
\end{array}
\right)  \label{43}
\end{equation}
We are looking for an appropriate gauge field such that the covariant
derivative becomes ${\not{D}}_A\rightarrow {\not{D}}+\not{A}$ with the
following gauge transformation rule:
\begin{equation}
\not{A}^{\prime }=G\not{A}G^{\dagger }-iG\left[ {\not{D}}_{dirac},G^{\dagger
}\right]  \label{44}
\end{equation}
A convenient choice is (see refs. \cite{15,16,17})
\begin{equation}
\not{A}=\left(
\begin{array}{cc}
iq\gamma ^\mu A_\mu ^{+} & \gamma ^5\Upsilon \\
\gamma ^5\overline{\Upsilon } & iq\gamma ^\mu A_\mu ^{-}
\end{array}
\right)  \label{45}
\end{equation}
where $\gamma ^\mu $ are the usual Dirac matrices and with
\begin{equation}
\left\{
\begin{array}{c}
\Upsilon =\varphi +\gamma ^5\phi \\
\overline{\Upsilon }=\varphi ^{*}-\gamma ^5\phi ^{*}
\end{array}
\right.  \label{46}
\end{equation}
where $\varphi $ and $\phi $ are the scalar components of the field $%
\Upsilon $. Those terms are fully equivalent to those in eq. (\ref{35bis}).
If $\Upsilon $ is different from zero, each charged particle of each brane
becomes sensitive to the electromagnetic fields of both branes irrespective
of their localization in the bulk. This kind of exotic interactions has been
considered previously in literature within the framework of mirror matter
\cite{25} and is not covered by the present paper. Moreover, to be
consistent with known physics, at least at low energies, $\Upsilon $ is
necessarily tiny (whereas $qA_\mu ^{\pm }$ needs not to be). This is
theoretically corroborated by eq. (\ref{44}) which shows that during each
gauge transformation, $\left| \varphi \right| $ (respectively $\left| \phi
\right| $) varies with an amplitude of order $g$ (respectively $\left|
m_c\right| $) whatever $\Lambda _{+}$ and $\Lambda _{-}$.

Using the covariant derivative ${\not{D}}_A\rightarrow {\not{D}}+\not{A}$
and according to expression (\ref{45}), the electromagnetic field can be
easily introduced in the two-brane Dirac equation (eq. (\ref{42})):
\begin{eqnarray}
\left( {i\not{D}}_A{-M}\right) \Psi &=&  \label{47} \\
&=&\left(
\begin{array}{cc}
i\gamma ^\mu (\partial _\mu +iqA_\mu ^{+})-m & ig\gamma ^5-m_c+i\gamma
^5\Upsilon \\
ig\gamma ^5-m_c^{*}+i\gamma ^5\overline{\Upsilon } & i\gamma ^\mu (\partial
_\mu +iqA_\mu ^{-})-m
\end{array}
\right) \left(
\begin{array}{c}
\psi _{+} \\
\psi _{-}
\end{array}
\right) =0  \nonumber
\end{eqnarray}

\subsection{Noncommutative two-sheeted spacetime vs two domain-walls}

It can immediately be noticed that eq. (\ref{35}) and eq. (\ref{47}) are
globally similar. In the two domain-walls approximation, $m\rightarrow
m+\delta m$ and $m_c=im_r$. The similarity between both equations was not
expected a priori and it suggests that eq. (\ref{47}) is quite generic in a
context of a two-level discretization of a 5D fermionic field. Note that in
the model introduced in refs. \cite{23,24}, $m_c=0$.

\begin{figure}[tbp]
\centerline{\ \includegraphics[width=14 cm]{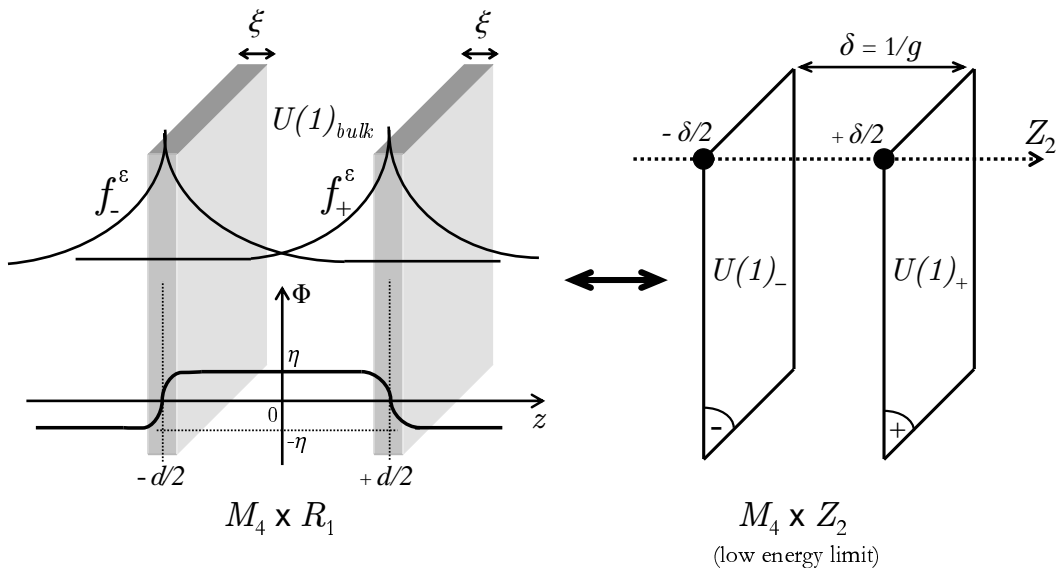}}
\caption{The two domain-walls in a $M_4\times R_1$ geometry are approximated
by a $M_4\times Z_2$ two-sheeted spacetime with an effective distance $%
\delta =\delta (\xi ,d,\varepsilon )$. The spatial extensions of the fermion
wave functions $f_{\pm }^\varepsilon $ depend on $\varepsilon $. The bulk $%
U(1)_{bulk}$ gauge group can be substituted by an effective $U(1)_{+}\otimes
U(1)_{-}$ gauge group where $U(1)_{+}$ (respectively $U(1)_{-}$) acts on the
brane $(+)$ (respectively $(-)$). In the paper, we simply refer to $%
U(1)_{bulk}$ as $U(1)$ and to $U(1)_{+}\otimes U(1)_{-}$ as $U(1)\otimes
U(1) $.}
\label{fig1}
\end{figure}

One notes that in the ''noncommutative'' two-sheeted approach, the term
proportional to $g$ (see (\ref{37})) is, by definition, a first order
discrete derivative along the $Z_2$ extra dimension. Similarly, in the
domain-wall approach, $g$ is related to the continuous extra dimension $R_1$
through an overlap integral (see section \ref{sec4} and appendix \ref
{appendixA}). From a pictorial perspective (see figure \ref{fig1}), the
two-brane world system ''collapses'' from a continuous description ($%
M_4\times R_1$) to a discrete one ($M_4\times Z_2$). Obviously the
similarity between eqs. (\ref{35}) and (\ref{47}) arises from the separation
ansatz made in eq. (\ref{24}) where the five-dimensional fermion is
restricted to the sum of two localized (i.e. four-dimensional) eigenmodes
with the same mass (see subsection \ref{sec3A}). Thus, by construction,
there are only two four-dimensional states in the model, and it seems that
the continuous extra dimension has been reduced to two points. As a
consequence, the continuous real extra dimension (and its continuous
derivative) is replaced by an effective phenomenological discrete extra
dimension (with its discrete derivative). The effective distance between
branes $\delta =1/g$ in noncommutative geometry is then related to the real
extra dimension through the integral in eqs. (\ref{33}) (see appendix \ref
{appendixB}).

One also notes that the five-dimensional $U(1)$ bulk gauge field is
substituted by an effective $U(1)\otimes U(1)$ gauge field acting in the $%
M_4\times Z_2$ spacetime (see figure \ref{fig1}). This is clearly
illustrated through the comparison between the gauge terms in eq. (\ref{35})
and in eq. (\ref{47}). This result could be expected a priori. Indeed, if
one considers a domain-wall at low energy, any valid gauge theory on the
bulk should lead to retrieve an effective $U(1)$ gauge field on the brane to
be consistent with known physics. As a consequence, for a set of two
domain-walls, it is not surprising to obtain an effective $U(1)\otimes U(1)$
theory. It is also interesting to note that the extra-dimensional component $%
\mathcal{A}_z$ of the bulk gauge field is related to the off-diagonal part $%
\Upsilon $ of the gauge field in the two-sheeted spacetime formalism (see
eqs. (\ref{35bis}) and (\ref{35ter})).

However, the link between the noncommutative two-sheeted spacetime and a
system of two domain-walls must be consider cautiously. The validity of such
a link rests on the following conditions:

$-$ As previously explained, eq. (\ref{35}) is derived for a single mass
state. We have just considered the lowest massive left-, right-states only,
i.e. the localized $n=1$ fermionic states (see subsection \ref{sec3A}). Of
course the $n>1$ modes could be used as well. Since we have neglected the
role of the heaviest fields, our model is therefore valid only for low
energies. From that point of view, the two-sheeted spacetime can be seen as
a simple low energy approximation of a two domain-walls system (see figure
\ref{fig1}). Nevertheless, a more general calculation would retain all mass
states and would be quite different from the presently considered
noncommutative model. Each domain-wall in the bulk would be then
approximated by a set of strongly coupled sheets, each one being related to
a specific mass instead of a single one. Moreover, excepted at high energy,
this would not change the main phenomelogical results of the present work
(hereafter discussed).

$-$ As mentioned in section \ref{sec4}, the two-level
approximation is valid as long as both branes are assumed to be
distant enough in the bulk. Indeed, the approximation assumes that
quadratic terms (and upper) implying overlap integrals can be
neglected (see appendices \ref{appendixA} and \ref {appendixB}).

\section{Non-relativistic limit and phenomenology}

As explained in the introduction, we are mainly concerned by low energies
phenomena occurring at a non-relativistic scale. To derive the
non-relativistic limit of the two-brane Dirac equation, we just observe that
eq. (\ref{47}) can also be written as:
\begin{equation}
\left(
\begin{array}{cc}
i\gamma ^\mu (\partial _\mu +iqA_\mu ^{+})-m & i\widetilde{g}\gamma ^5-%
\widetilde{m}_c \\
i\widetilde{g}^{*}\gamma ^5-\widetilde{m}_c^{*} & i\gamma ^\mu (\partial
_\mu +iqA_\mu ^{-})-m
\end{array}
\right) \left(
\begin{array}{c}
\psi _{+} \\
\psi _{-}
\end{array}
\right) =0  \label{47a}
\end{equation}
with
\begin{equation}
\left\{
\begin{array}{c}
\widetilde{g}=g+\varphi  \\
\widetilde{m}_c=m_c-i\phi
\end{array}
\right.   \label{47bis}
\end{equation}
We have just replaced the field $\Upsilon $ and the coupling parameters $g$
and $m_c$ by the effective fields $\widetilde{g}$ and $\widetilde{m}_c$ as
shown by eq. (\ref{47bis}). Without loss of generality, we will consider now
that $\widetilde{g}\approx g$ and $\widetilde{m}_c\approx m_c$ since $\left|
\varphi \right| $ (respectively $\left| \phi \right| $) should not exceed $g$
(respectively $\left| m_c\right| $) as explained before. This choice allows
to further simplify the model. It is somewhat equivalent to set the
off-diagonal term $\Upsilon $ to zero. With such a choice, we simply assume
that the electromagnetic field of a brane couples only with the particles
belonging to the same brane. Each brane possesses its own current and charge
density distribution as sources of the local electromagnetic fields. On the
two branes live then the distinct $A_\mu ^{+}$ and $A_\mu ^{-}$
electromagnetic fields. The photon fields $A_\mu ^{\pm }$ behave
independently from each other and are totally trapped in their original
brane in accordance with observations: photons belonging to a given brane
are not able to reach the other brane. As a noticeable consequence, the
structures belonging to the branes are mutually invisible by local
observers. Without loss of generality, eq. (\ref{47a}) can be recast as:
\begin{equation}
\left(
\begin{array}{cc}
i\gamma ^\mu (\partial _\mu +iqA_\mu ^{+})-m & ig\gamma ^5-m_c \\
ig\gamma ^5-m_c^{*} & i\gamma ^\mu (\partial _\mu +iqA_\mu ^{-})-m
\end{array}
\right) \left(
\begin{array}{c}
\psi _{+} \\
\psi _{-}
\end{array}
\right) =0  \label{47ter}
\end{equation}

Following the well-known standard procedure, a two-brane Pauli equation can
then be derived:
\begin{equation}
i\hbar \frac \partial {\partial t}\left(
\begin{array}{c}
\psi _{+} \\
\psi _{-}
\end{array}
\right) =\left\{ \mathbf{H}_0+\mathbf{H}_{cm}+\mathbf{H}_c+\mathbf{H}%
_s\right\} \left(
\begin{array}{c}
\psi _{+} \\
\psi _{-}
\end{array}
\right)  \label{48}
\end{equation}
where $\psi _{+}$ and $\psi _{-}$ correspond to the wave functions in the $%
(+)$ and $(-)$ branes respectively. $\psi _{+}$ and $\psi _{-}$ are here
Pauli spinors. The Hamiltonian $\mathbf{H}_0$ is a block-diagonal matrix
\begin{equation}
\mathbf{H}_0=\left(
\begin{array}{cc}
\mathbf{H}_{+} & 0 \\
0 & \mathbf{H}_{-}
\end{array}
\right)  \label{49}
\end{equation}
where each block is simply the classical Pauli Hamiltonian expressed in each
branes:
\begin{equation}
\mathbf{H}_{\pm }=-\frac{\hbar ^2}{2m}\left( \mathbf{\nabla }-i\frac q\hbar
\mathbf{A}_{\pm }\right) ^2+g_s\mu \frac 12\mathbf{\sigma \cdot B}_{\pm
}+V_{\pm }  \label{50}
\end{equation}
such that $\mathbf{A}_{+}$ and $\mathbf{A}_{-}$ correspond to the magnetic
vector potentials in the branes $(+)$ and $(-)$ respectively. The same
convention is applied to the magnetic fields $\mathbf{B}_{\pm }$ and to the
potentials $V_{\pm }$. $g_s\mu $ is the magnetic moment of the particle with
$g_s$ the gyromagnetic factor and $\mu $ the magneton. In addition to these
``classical'' terms, the two-brane Hamiltonian comprises also new terms
specific of the two-brane world:
\begin{eqnarray}
\mathbf{H}_{cm}=igg_s\mu \frac 12\left(
\begin{array}{cc}
0 & -\mathbf{\sigma \cdot }\left\{ \mathbf{A}_{+}-\mathbf{A}_{-}\right\} \\
\mathbf{\sigma \cdot }\left\{ \mathbf{A}_{+}-\mathbf{A}_{-}\right\} & 0
\end{array}
\right)  \label{51}
\end{eqnarray}
and
\begin{equation}
\mathbf{H}_c=\left(
\begin{array}{cc}
0 & m_cc^2 \\
m_c^{*}c^2 & 0
\end{array}
\right)  \label{52}
\end{equation}
and
\begin{equation}
\mathbf{H}_s=\frac{g^2\hbar ^2}{2m}\left(
\begin{array}{cc}
1 & 0 \\
0 & 1
\end{array}
\right)  \label{53}
\end{equation}

It can be noticed that $\mathbf{H}_s$ is constant and vanishes through a
convenient energy rescaling. By contrast, $\mathbf{H}_c$ and $\mathbf{H}%
_{cm} $ are non conventional Hamiltonian components whose effects are now
discussed.

\subsection{Spontaneous oscillations between branes?}

$\mathbf{H}_c$ is a constant term resulting from the off-diagonal mass terms
(eq. (\ref{41})). It can be responsible for free spontaneous oscillations
between the two branes. Let us illustrate this. For sake of simplicity we
assume that $\mathbf{B}_{\pm }=\mathbf{0}$ and $\mathbf{A}_{\pm }=\mathbf{0}%
. $ $\mathbf{H}_{\pm }$ reduces then to $\mathbf{H}_{\pm }=V_{\pm }$ and $%
\mathbf{H}_{cm}=0$. From eq. (\ref{48}), it can be shown that a particle
initially ($t=0$) localized in our brane will have a probability to be
located in the other brane at time $t$ given by:
\begin{equation}
P(t)=\frac{4\Omega _c^2}{\Omega _0^2+4\Omega _c^2}\sin ^2\left( (1/2)\sqrt{%
\Omega _0^2+4\Omega _c^2}t\right)  \label{54}
\end{equation}
where $\Omega _c=\left| m_c\right| c^2/\hbar $ and $\Omega
_0=(V_{+}-V_{-})/\hbar $. Eq. (\ref{54}) shows that the particle
undergoes Rabi-like oscillations between both branes. $\Omega
_0\hbar $ is an effective potential mimicking the interactions of
the particle with its environment \cite{23}. $\Omega _0\hbar $
might contain the contribution of atomic nuclei electrostatic
fields or the Earth's gravitational field for instance \cite{23}.
An important point is that the oscillations are strongly
suppressed when $\Omega _0\hbar $ becomes greater than $\left|
m_c\right| c^2 $, i.e. when the particle is strongly interacting
with its environment through $\mathbf{H}_0$. As a consequence,
these spontaneous oscillations will probably be hardly observed.

\subsection{Induced matter swapping between branes}

\label{subsec72}

$\mathbf{H}_{cm}$ is a geometrical coupling involving electromagnetic fields
of the two branes. $\mathbf{H}_{cm}$ vanishes for null magnetic vector
potentials. $\mathbf{H}_{cm}$ also implies Rabi-like oscillations between
the branes. This effect was previously considered in previous papers in
which the reader will find more detailed explanations \cite{23}. It is only
important here to remind that eq. (\ref{48}) holds resonant solutions for a
magnetic vector potential rotating with an angular frequency $\omega $. One
may consider for instance a neutral particle, endowed with a magnetic
moment, initially ($t=0$) localized in our brane in a region of curlless
rotating magnetic vector potential such that $\mathbf{A}_{-}=A_p\mathbf{e(}t%
\mathbf{)}$ and $\mathbf{A}_{+}=0$, with $\mathbf{e}(t)=\left( \cos \omega
t,\sin \omega t,0\right) $. $\omega $ is the angular frequency of the field $%
\mathbf{A}_{-}$ and can be possibly null (static field case). Let us assume
that the conventional part of the Pauli Hamiltonian $\mathbf{H}_{\pm }$ can
be written as $\mathbf{H}_{\pm }=V_{\pm }$. Moreover, $\mathbf{H}_c$ is
neglected relative to $\mathbf{H}_{cm}$. From eq. (\ref{48}), it can be
shown that any particle initially in a spin-down state for instance
(according to $\mathbf{e}_3=\left( 0,0,1\right) $) and localized in our
brane at $t=0$ can be detected in the second brane at time $t$ with a
probability \cite{23}:
\begin{equation}
P(t)=\frac{4\Omega _p^2}{(\Omega _0-\omega )^2+4\Omega _p^2}\sin ^2\left(
(1/2)\sqrt{(\Omega _0-\omega )^2+4\Omega _p^2}t\right)  \label{55}
\end{equation}
where $\Omega _p=gg_s\mu A_p/(2\hbar )$ and $\Omega _0=(V_{+}-V_{-})/\hbar $%
. In addition, in the second brane, the particle is then in a spin-up state.
$\Omega _0\hbar $ is still an effective potential mimicking the interactions
of the particle with its environment \cite{23}. Eq. (\ref{55}) shows how the
induced matter swapping between branes occurs through a Rabi-like
phenomenon. The resonant exchange occurs whenever the magnetic vector
potential rotates with an angular frequency $\omega =\Omega _0$. It is clear
that the situation described in eq. (\ref{55}) remains rather simplistic.
More realistic descriptions to achieve experimental conditions of matter
swapping are suggested elsewhere \cite{26}. It can be easily checked that a
static field case allows matter swapping to occur as well. However, the
amplitude of the vector potential must be huge to overcome the particle
confinement induced by the environment \cite{23}. From an experimental point
of view, the resonant mechanism appears as a worth studying alternative.

Again, it is necessary to stress that the matter swapping mechanism
described here depends on the $\mathbf{H}_{cm}$ term only. Since this term
is predicted by distinct mathematical approaches, we expect that induced
matter swapping between branes might be a generic phenomenon of any 5D model
containing two lower-dimensional (4D) sheets.

\section{Discussion and conclusions}

$\qquad\bullet $ Rewriting the 5D continuous Dirac equation in a two-level
form presents many advantages:

$-$ It allows a dramatic simplification of the equations and allows a better
understanding of the quantum behavior of particles in a two-brane world
setup.

$-$ Connes et al have shown the great potential of noncommutative geometries
\cite{14}, especially of two-sheeted spacetime representations which are
suitable to recover the Standard Model of particles \cite{14,15}. In the
present paper, a formal equivalence between domain-walls approaches and
certain noncommutative geometries has been shown. ''Non-commutativity''
appears here as a consequence of a two-level simplification. Nevertheless,
the arena where physical events take place still remains commutative. This
unexpected bridge between domain-walls and non-commutativity clearly
deserves further studies.\\

$\bullet $ In the present model, the kink-domain-wall localized at $z=-d/2$
(i.e. the $(-)$ brane) undergoes left-handed neutrinos while the
antikink-domain-wall localized at $z=d/2$ (i.e. the $(+)$ brane) undergoes
right-handed neutrinos. As a consequence, and due to the doubling of the
wave function, this can be seen as a reminiscence of the mirror-matter
concept \cite{27}. Nevertheless, while it is true that hidden sector models
and present approach share several common points, it is equally true that
they differ in many ways:

$-$ In the mirror matter formalism, there is only one 4D manifold which
justifies for the left/right parity by introducing implicit new internal
degrees of freedom to particles. In the present work, it can be noted that
the number of particle families remains unchanged but the particles have now
access granted to two distinct branes.

$-$ Moreover, in the mirror matter approach, the mixing between our world
and mirror-world occurs through a photon/mirror photon kinetic mixing \cite
{25} (gravitation also is assumed to mediate interactions between matter and
mirror matter but it is not relevant in the present discussion).
Nevertheless, the present two-brane structure demonstrates the existence of
oscillations without recourse to a photon/mirror photon kinetic mixing.\\

$\bullet $ It is suggested that the ''swapping effect'' (see subsection \ref
{subsec72}) involving induced Rabi-like oscillations of matter between
adjacent branes might be a common feature of any braneworld theory involving
more than one brane in the bulk. Indeed, the method used in section \ref
{sec4} to obtain the relevant eq. (\ref{35}) is quite general and it does
not rely on any assumption concerning the domain-walls. The Hamiltonian term
$\mathbf{H}_{cm}$ (see eq. (\ref{51})) is due to the $ig\gamma ^5$ terms in
eq. (\ref{47ter}), which are related to the overlap integral over the fifth
dimension of the product of the extra-dimensional fermionic wave functions
related to each brane. We expect that in more complex domain-walls models
(involving warped metric for instance, like in Randall-Sundrum brane
worlds), we would just have obtained different expressions for $g$, $m_r$
and $\delta m$ (see eqs. (\ref{33})).

Finally, if one considers two parallel 3-branes in a bulk with more than
five dimensions (say $3+N+1$, with $N>1$), the situation should be quite
similar to that described in this paper. Indeed, considering a bijective
relation between the two 3-branes, one can build a fibre bundle linking each
point of the branes. Each fibre allows to define a preferential fifth
dimension connecting both branes. Moreover, since the fermionic wave
functions spread over the $N-1$ other extra dimensions, they must quickly
decrease when going away from the branes. The system would therefore reduce
from $3+N+1$ to $3+1+1$ dimensions similar to the setup considered in this
paper \cite{9}.

As a consequence, we conjecture that at low energy, any multidimensional
setup containing two branes can be described by a two-sheeted spacetime in
the formalism of the noncommutative geometry. As a result, we also
conjecture that the so-called ''swapping effect'' originally predicted in
the context of $M_4\times Z_2$ geometries \cite{23,24} might be a
model-independent feature of any multidimensional setup containing at least
two branes. Note that, due to the links expected between domain-walls and
string theories \cite{21}, one might also wonder to what extent the
''swapping effect'' described here could be hidden in string theories as
well.

\appendix

\section{Derivation of the explicit expression of $\widetilde{H}$}

\label{appendixA}

In the following, we detail how eq. (\ref{32}) can be derived from eq. (\ref
{31}). In eq. (\ref{31}), the two-level Hamiltonian $\widetilde{H}$ was
written as
\begin{equation}
\widetilde{H}=\frac 1{1-s^2}\otimes \left(
\begin{array}{cc}
h_{+,+}-sh_{-,+} & h_{+,-}-sh_{-,-} \\
h_{-,+}-sh_{++} & h_{-,-}-sh_{+,-}
\end{array}
\right)  \label{A1}
\end{equation}
Using equations (\ref{6}), (\ref{10}), (\ref{11}), (\ref{14})-(\ref{18}), (%
\ref{26}) and (\ref{27}), the terms of (\ref{A1}) can be expressed as
follows
\begin{equation}
s=a+\gamma ^5b  \label{A2}
\end{equation}
with
\begin{eqnarray}
\left\{
\begin{array}{c}
a=\int \left\{ f_{+}(z)f_{-}(z)+\kappa _{-}(z)\kappa _{+}(z)\right\} dz \\
b=\int \left\{ f_{-}(z)\kappa _{+}(z)+\kappa _{-}(z)f_{+}(z)\right\} dz
\end{array}
\right.  \label{A3}
\end{eqnarray}
and
\begin{equation}
h_{i,j}=\left\{ \int \Pi _i^{\dagger }(z)\Pi _j(z)dz\right\}
\left\{ -i\gamma ^0\gamma ^\eta \partial _\eta \right\} +\gamma
^0\gamma ^5\alpha _{i,j}+\gamma ^0\beta _{i,j}+\gamma ^0
G_{i,j}+\gamma ^0\gamma ^5 \widetilde{G}_{i,j} \label{A4}
\end{equation}
with
\begin{equation}
\left\{
\begin{array}{c}
\alpha _{i,j}=\lambda \int \left\{ \Phi _i(z)+\Delta \Phi \right\} \left\{
f_i(z)\kappa _j(z)-\kappa _i(z)f_j(z)\right\} dz-mb\text{ if }i\neq j \\
\alpha _{i,i}=0
\end{array}
\right.  \label{A5}
\end{equation}
and
\begin{equation}
\left\{
\begin{array}{c}
\beta _{i,j}=\lambda \int \left\{ \Phi _i(z)+\Delta \Phi \right\} \left\{
f_i(z)f_j(z)-\kappa _i(z)\kappa _j(z)\right\} dz+ma\text{ if }i\neq j \\
\beta _{i,i}=m+\lambda \int \Phi _j(z)\left\{ f_i^2(z)-\kappa
_i^2(z)\right\} dz\text{ where }i\neq j
\end{array}
\right.  \label{A6}
\end{equation}
We note that, according to eq. (\ref{5}) and eq. (\ref{6}), $\Phi _{\pm
}(z)=\mp \Phi (z\mp d/2)$ and $\Delta \Phi =-\eta $. $\alpha _{i,j}$ is
related to the projection of the first derivative $\partial _z$ term of $H$
(see eq. (\ref{26})) onto the independent states of each brane. $\beta
_{i,j} $ is related to the projection of the scalar field $\Phi $ term of $H$
onto the independent states of each brane.

Recalling that $\mathcal{A}(x,z)=\mathcal{A}^{+}(x,z)+\mathcal{A}^{-}(x,z)$
(see subsection \ref{TBGF}) and using $\mathcal{\not{A}}{=}\gamma ^\mu
\mathcal{A}_\mu $, one also gets:
\begin{equation}
\left\{
\begin{array}{c}
G_{i,j}=\mathcal{\not{A}}_{i,j}^{+}+\mathcal{\not{A}}_{i,j}^{-}-i\mathcal{A}%
_{z,i,j}^{+}-i\mathcal{A}_{z,i,j}^{-} \\
\widetilde{G}_{i,j}=\mathcal{\not{B}}_{i,j}^{+}+\mathcal{\not{B}}_{i,j}^{-}+i%
\mathcal{B}_{z,i,j}^{+}+i\mathcal{B}_{z,i,j}^{-}
\end{array}
\right.   \label{A7}
\end{equation}
with
\begin{eqnarray}
\left\{
\begin{array}{c}
\mathcal{\not{A}}_{i,j}^{\pm }=\int \left\{ f_i(z)f_j(z)+\kappa _i(z)\kappa
_j(z)\right\} \mathcal{\not{A}}^{\pm }dz \\
\mathcal{A}_{z,i,j}^{\pm }=\int \left\{ f_i(z)\kappa _j(z)-\kappa
_i(z)f_j(z)\right\} \mathcal{A}_z^{\pm }dz \\
\mathcal{\not{B}}_{i,j}^{\pm }=-\int \left\{ f_i(z)\kappa _j(z)+\kappa
_i(z)f_j(z)\right\} \mathcal{\not{A}}^{\pm }dz \\
\mathcal{B}_{z,i,j}^{\pm }=-\int \left\{ f_i(z)f_j(z)-\kappa _i(z)\kappa
_j(z)\right\} \mathcal{A}_z^{\pm }dz
\end{array}
\right.   \label{A8}
\end{eqnarray}
Hence, the Hamiltonian (\ref{A1}) can be written as:
\begin{eqnarray}
\widetilde{H} &=&\left(
\begin{array}{cc}
-i\gamma ^0\gamma ^\eta \partial _\eta  & 0 \\
0 & -i\gamma ^0\gamma ^\eta \partial _\eta
\end{array}
\right) +  \label{A9} \\
&&\frac 1{1-s^2}\otimes \left(
\begin{array}{c}
\gamma ^0\left( \beta _{+,+}+b\alpha _{-,+}-a\beta _{-,+}\right) -\gamma
^0\gamma ^5\left( a\alpha _{-,+}-b\beta _{-,+}\right)  \\
\gamma ^0\gamma ^5\left( \alpha _{-,+}+b\beta _{+,+}\right) +\gamma ^0\left(
\beta _{-,+}-a\beta _{+,+}\right)
\end{array}
\right.   \nonumber \\
&&\left.
\begin{array}{c}
\gamma ^0\gamma ^5\left( \alpha _{+,-}+b\beta _{-,-}\right) +\gamma ^0\left(
\beta _{+,-}-a\beta _{-,-}\right)  \\
\gamma ^0\left( \beta _{-,-}+b\alpha _{+,-}-a\beta _{+,-}\right) -\gamma
^0\gamma ^5\left( a\alpha _{+,-}-b\beta _{+,-}\right)
\end{array}
\right) +  \nonumber \\
&&\frac 1{1-s^2}\otimes \left(
\begin{array}{c}
\gamma ^0\left( G_{+,+}-aG_{-,+}-b\widetilde{G}_{-,+}\right) +\gamma
^0\gamma ^5\left( \widetilde{G}_{+,+}-a\widetilde{G}_{-,+}-bG_{-,+}\right)
\\
\gamma ^0\left( G_{-,+}-aG_{+,+}-b\widetilde{G}_{+,+}\right) +\gamma
^0\gamma ^5\left( \widetilde{G}_{-,+}-a\widetilde{G}_{+,+}-bG_{+,+}\right)
\end{array}
\right.   \nonumber \\
&&\left.
\begin{array}{c}
\gamma ^0\left( G_{+,-}-aG_{-,-}-b\widetilde{G}_{-,-}\right) +\gamma
^0\gamma ^5\left( \widetilde{G}_{+,-}-a\widetilde{G}_{-,-}-bG_{-,-}\right)
\\
\gamma ^0\left( G_{-,-}-aG_{+,-}-b\widetilde{G}_{+,-}\right) +\gamma
^0\gamma ^5\left( \widetilde{G}_{-,-}-a\widetilde{G}_{+,-}-bG_{+,-}\right)
\end{array}
\right)   \nonumber
\end{eqnarray}

The integrals $a$, $b$, $\alpha _{i,j}$ (with $i\neq j$) and $\beta _{i,j}$
(with $i\neq j$) involve the overlapping of the fermionic wave functions of
each brane. These terms have to be small enough to act as a correction and
for a sake of simplicity, we assume that higher-order terms (i.e. $a^2$, $%
b^2 $, $ab$, $a\alpha _{\mp ,\pm }$, $b\alpha _{\mp ,\pm }$, $a\beta _{\mp
,\pm } $, $b\beta _{\mp ,\pm }$, ...) can be fairly neglected.

The terms related to the gauge field (see eqs. (\ref{A7}) and (\ref{A8}))
must be considered with caution. Indeed, since the gauge terms are related
to quantum corrections of the 5D gauge field, one must take care of
undesirable anomalies that could break the gauge invariance or the
hermiticity of the two-level Hamiltonian. Assuming for instance that the
Hamiltonian remains Hermitic requires these undesirable terms to vanish. In
practice, the terms inducing anomalies can be cancelled by adding convenient
extra fermion species, or compensated through some quantum number flows in
the bulk \cite{28a}. In addition, the integrals $G_{i,j}$ and $\widetilde{G}%
_{i,j}$ (with $i\neq j$) also imply an overlapping of the fermionic wave
functions of each brane. These terms should act as perturbative terms. For
similar reasons, the higher-order terms $aG_{i,j}$, $bG_{i,j}$, $aG_{i,j}$
and $bG_{i,j}$ can be fairly neglected. Keeping the only relevant gauge
field terms (see section \ref{GF}), the Hamiltonian (\ref{A9}) reduces then
to:
\begin{equation}
\widetilde{H}=\left(
\begin{array}{cc}
-i\gamma ^0\gamma ^\eta \partial _\eta +\gamma ^0m+\gamma ^0\delta
m_{+}+\gamma ^0\mathcal{\not{A}}_{+,+}^{+} & \gamma ^0\gamma
^5g_{+,-}+\gamma ^0m_{+,-}-\gamma ^0\gamma ^5\Upsilon  \\
\gamma ^0\gamma ^5g_{-,+}+\gamma ^0m_{-,+}+\gamma ^0\gamma ^5\overline{%
\Upsilon } & -i\gamma ^0\gamma ^\eta \partial _\eta +\gamma ^0m+\gamma
^0\delta m_{-}+\gamma ^0\mathcal{\not{A}}_{-,-}^{-}
\end{array}
\right)   \label{A10}
\end{equation}
where
\begin{eqnarray}
\left\{
\begin{array}{c}
g_{\pm ,\mp }=\mp \lambda \int \left\{ \Phi _{\pm }+\Delta \Phi \right\}
\left\{ f_{-}(z)\kappa _{+}(z)-\kappa _{-}(z)f_{+}(z)\right\} dz \\
m_{\pm ,\mp }=\lambda \int \left\{ \Phi _{\pm }+\Delta \Phi \right\} \left\{
f_{-}(z)f_{+}(z)-\kappa _{-}(z)\kappa _{+}(z)\right\} dz \\
\delta m_{\pm }=\lambda \int \Phi _{\mp }\left\{ f_{\pm }^2(z)-\kappa _{\pm
}^2(z)\right\} dz
\end{array}
\right.   \label{A11}
\end{eqnarray}
and
\begin{equation}
\left\{
\begin{array}{c}
\Upsilon =\varphi +\gamma ^5\phi  \\
\overline{\Upsilon }=\varphi ^{*}-\gamma ^5\phi ^{*}
\end{array}
\right.   \label{A12}
\end{equation}
with
\begin{equation}
\left\{
\begin{array}{c}
\phi =i\int \left\{ f_{+}(z)\kappa _{-}(z)-\kappa _{+}(z)f_{-}(z)\right\}
\left\{ \mathcal{A}_z^{+}+\mathcal{A}_z^{-}\right\} dz \\
\varphi =i\int \left\{ f_{+}(z)f_{-}(z)-\kappa _{+}(z)\kappa _{-}(z)\right\}
\left\{ \mathcal{A}_z^{+}+\mathcal{A}_z^{-}\right\} dz
\end{array}
\right.   \label{A13}
\end{equation}
Since $\widetilde{H}$ must be hermitic, the properties of $g_{ij}$ and $%
m_{ij}$ can be easily deduced. We get $g_{+,-}=-g_{-,+}=-g$ and $%
m_{-,+}=m_{+,-}=m_r$. Due to the symmetry between both branes, we get $%
\delta m_{\pm }=\delta m$. Finally the effective Hamiltonian reads:
\begin{equation}
\widetilde{H}=\left(
\begin{array}{cc}
-i\gamma ^0\gamma ^\eta (\partial _\eta +iqA_\eta ^{+})+\gamma ^0m+\gamma
^0\delta m+qA_0^{+} & -\gamma ^0\gamma ^5g+\gamma ^0m_r-\gamma ^0\gamma
^5\Upsilon  \\
\gamma ^0\gamma ^5g+\gamma ^0m_r+\gamma ^0\gamma ^5\overline{\Upsilon } &
-i\gamma ^0\gamma ^\eta (\partial _\eta +iqA_\eta ^{-})+\gamma ^0m+\gamma
^0\delta m+qA_0^{-}
\end{array}
\right)   \label{A14}
\end{equation}
where we have assumed that
\begin{equation}
qA_\mu ^{\pm }=\int \left\{ f_{\pm }^2(z)+\kappa _{\pm }^2(z)\right\}
\mathcal{A}_\mu ^{\pm }(x,z)dz  \label{A15}
\end{equation}
with the field redefinition $a_\mu ^{\pm }\rightarrow qA_\mu ^{\pm }$. The
constant with the dimension of a charge $q$ gives to the effective gauge
vector fields the correct usual physical dimensions. $A_\mu ^{\pm }$
correspond to the usual electromagnetic fields onto the brane $(+)$ or $(-)$.

Assuming the first order approximation, $g$ (respectively $m_r$) is only
related to $\alpha _{i,j}$ (respectively $\beta _{i,j}$) (see eq. (\ref{A5})
and eq. (\ref{A6})). In that case, $g$ depends only on the first order
derivative $\partial _z$ along the continuous extra dimension while $m_r$ is
only related to the scalar field $\Phi $.

\section{Estimation of $g$, $m_r$ and $\delta m$}

\label{appendixB} In the present appendix we consider the behavior of $g$, $%
m_r$ and $\delta m$ from the expressions given by eqs. (\ref{33}). Using
expressions (\ref{6}) and (\ref{18}), eqs. (\ref{33}) can be written as:

\begin{eqnarray}
\left\{
\begin{array}{c}
g=(1/2)\lambda \eta \int \left\{ \tanh \left( \left( z+d/2\right) /\xi
\right) -1\right\} \left\{ f_{R,+}(z)f_{L,-}(z)-f_{L,+}(z)f_{R,-}(z)\right\}
dz \\
m_r=(1/2)\lambda \eta \int \left\{ \tanh \left( \left( z+d/2\right) /\xi
\right) -1\right\} \left\{ f_{L,+}(z)f_{R,-}(z)+f_{R,+}(z)f_{L,-}(z)\right\}
dz \\
\delta m=\lambda \eta \int \tanh \left( \left( z+d/2\right) /\xi \right)
f_{R,+}(z)f_{L,+}(z)dz
\end{array}
\right.  \label{B1}
\end{eqnarray}
The expressions of $f_{L/R,\pm }(z)$ are easily deduced from expressions (%
\ref{21}) to (\ref{23}). It can be shown (with the help of a numerical tool
like \textit{Mathematica}$^{\textregistered}$ for instance) that there is no
trivial analytical expressions for such integrals (\ref{B1}) except if $%
\varepsilon =\lambda \sqrt{2/\chi }=\lambda \eta \xi $ is an integer. In
that case, the above integrals can be written as a ratio of two functions.
Each function appears then as a sum of exponential terms: $\exp (-nd/\xi )$
and $(d/\xi )\exp (-nd/\xi )$ where $n$ are integers. In the two-brane
solutions considered here, we have $d\gg \xi $ and the integrals can then be
approximated by a single exponential function. By considering the first
massive mode, several cases can be considered related to the mass $m=m_1=k%
\sqrt{2\varepsilon -1}$ of the particle trapped on a brane (we set $k=1/\xi $%
):

\[
\begin{tabular}{|c|c|c|c|c|}
\hline
\  & Fermion mass $m$ & Coupling Constant $g$ & Off-Diagonal Mass $m_r$ &
Mass Correction $\delta m$ \\[1ex] \hline
$\varepsilon =1$ & $k$ & $0$ & $0$ & $0$ \\[1ex] \hline
$\varepsilon =2$ & $k\sqrt{3}$ & $2ke^{-kd}$ & $-2ke^{-kd}$ & $8\sqrt{3}%
k(kd-1)e^{-2kd}$ \\[1ex] \hline
$\varepsilon =3$ & $k\sqrt{5}$ & $24ke^{-2kd}$ & $6ke^{-2kd}$ & $6\sqrt{5}%
ke^{-2kd}$ \\[1ex] \hline
$\varepsilon =4$ & $k\sqrt{7}$ & $180ke^{-3kd}$ & $100ke^{-3kd}$ & $4\sqrt{7}%
ke^{-2kd}$ \\[1ex] \hline
$\varepsilon =5$ & $3k$ & $1120ke^{-4kd}$ & $770ke^{-4kd}$ & $10ke^{-2kd}$ \\%
[1ex] \hline
\end{tabular}
\]

A first noticeable point is that the coupling constant $g$ decreases when $%
\varepsilon $ increases. Indeed, it should be kept in mind that the
fermionic wave functions become sharply localized when $\varepsilon $
increases (see subsection \ref{sec3A}), such that the overlap of the
fermionic states of each brane also decreases. It is also noticeable that $g$
can be written as $g=f_\varepsilon k\exp (-nkd)$ (where $n$ is an integer
related to $\varepsilon $ and $f_\varepsilon $ a constant which depends of $%
\varepsilon $). As a consequence, the phenomenological distance $\delta $
between the branes becomes $\delta =f_\varepsilon ^{-1}\xi \exp (nd/\xi )$
and is related to the real distance $d$ between branes and to the brane
thickness $\xi $.

Let us consider, for instance, the case $\varepsilon =2$ (for which $m=k%
\sqrt{3}=\sqrt{3}/\xi $) assuming a mass $m$ equals to the electron mass. We
get $g=10^3$ m$^{-1}$ i.e. $\delta =1$ mm when $d=1,46\cdot 10^{-11}$ m
(i.e. about $22$ times the brane thickness). For $d=2,38\cdot 10^{-11}$ m
(i.e. about $36$ times the brane thickness) we get $g=10^{-3}$ m$^{-1}$ i.e.
$\delta =1$ km. Therefore, the real distance $d$ between branes is in
agreement with our $d\gg \xi $ hypothesis. Furthermore, the values of the
coupling constant $g$ are also consistent with observable phenomena in the
context of present day technology \cite{23,24,26,28}. It can be noticed that
slight fluctuations of the actual distance $d $ between branes lead to
strong variations of the coupling constant $g$ ($g$ is multiplied by $10^6$
when $d$ is divided by $1,6$). Finally, for a range of distances $d$ between
branes, the model remains fully consistent with constraints induced by
Newton law variation measurements \cite{13}.

\end{document}